\documentclass[fleqn,twoside]{article}
\usepackage{espcrc2}

\usepackage{graphicx}
\usepackage[figuresright]{rotating}

\newcommand{\AmS}{{\protect\the\textfont2
  A\kern-.1667em\lower.5ex\hbox{M}\kern-.125emS}}

\title{The measurement of $R$ at CLEO}

\author{J. Libby\address[OXF]{University of Oxford, Denys Wilkinson Building, Keble Road, 
Oxford, OX1 3RH, United Kingdom} (on behalf of the CLEO collaboration)}
       
\begin{document}

\begin{abstract}
Measurements of the total cross section of $e^{+}e^{-}\rightarrow\mathrm{hadrons}$ are 
presented in two different ranges of centre-of-mass energy. The measurements are made using 
the CLEO III and CLEO-c detectors at the Cornell Electron Storage Ring. The absolute cross 
sections and the values of $R$, the ratio of hadronic to muon pair production cross 
sections, are determined at seven centre-of-mass energies between 6.964 and 10.538 GeV. The 
total cross sections and values of $R$ are also determined at thirteen centre-of-mass 
energies between 3.97 and 4.26~GeV; in addition, the inclusive and exclusive cross sections 
for $D^{+}$, $D^{0}$ and $D^{+}_{s}$ production are presented. Furthermore, for the lower 
centre-of-mass energy range, exclusive cross-sections are presented for final states 
consisting of two charm mesons: $D\overline{D}$, $D^{*}\overline{D}$, $D\overline{D^{*}}$, 
$D^{*}\overline{D^{*}}$, $D^{+}_{s}D^{-}_{s}$, $D^{*+}_{s}D^{-}_{s}$,
$D^{+}_{s}D^{*-}_{s}$, $D\overline{D}^{*}\pi$ and $D^{*}\overline{D}^{*}\pi$.    
\vspace{1pc}
\end{abstract}

\maketitle

\section{Introduction}
The determination of $R$, the ratio of the radiation-corrected hadronic cross 
section to the calculated lowest-order cross section for muon pair production is presented in 
two ranges of centre-of-mass energy, $\sqrt{s}$. The motivation for these studies is 
different in the two $\sqrt{s}$ ranges. 

The measurements of $R$ at seven values of $\sqrt{s}$ between 6.964 and 10.538~GeV test the 
asymptotic freedom of the QCD coupling $\alpha_{s}$ in the range of $\sqrt{s}$ where $u$, 
$d$, $s$ and $c$, quarks are produced. The values of $\alpha_{s}$ are extracted by comparing 
the measurements of $R$ to a perturbative QCD calculation at the four-loop level 
\cite{bib:alphas_theory}.  

Measurements of $R$ in the region just above the $c\overline{c}$ threshold ($\sqrt{s}=3.970$ 
to 4.260~GeV) exhibit a rich structure (for example see Ref. \cite{bib:bes_R}); this reflects 
the production of $c\overline{c}$ resonances. Interesting features include an enhancement at 
$D^{*}\overline{D^{*}}$ threshold $(\sqrt{s}=4.02~GeV)$ and a broad plateau beginning at the 
$D^{*+}_{s}D^{-}_{s}$ threshold $(\sqrt{s}=4.08~GeV)$. There is considerable theoretical 
interest in the composition of these enhancements \cite{bib:ccbar_threshold_theory}. However, 
there is limited experimental information available which motivates the exclusive cross 
sections presented in this paper. The following final states are considered: $D\overline{D}$, 
$D^{*}\overline{D}$, $D\overline{D^{*}}$, $D^{*}\overline{D^{*}}$, $D^{+}_{s}D^{-}_{s}$, 
$D^{*+}_{s}D^{-}_{s}$ and $D^{+}_{s}D^{*-}_{s}$, $D\overline{D}^{*}\pi$ and 
$D^{*}\overline{D}^{*}\pi$. 
The total cross sections and values of $R$ are also determined at the thirteen points 
studies. 


The results given in this paper are presented in greater detail elsewhere 
\cite{bib:cleoIII_R,bib:cleoc_R}. 

\section{CLEO III and CLEO-c}
\label{sec:cleo}

\begin{figure}[htb]
\includegraphics*[width=0.95\columnwidth]{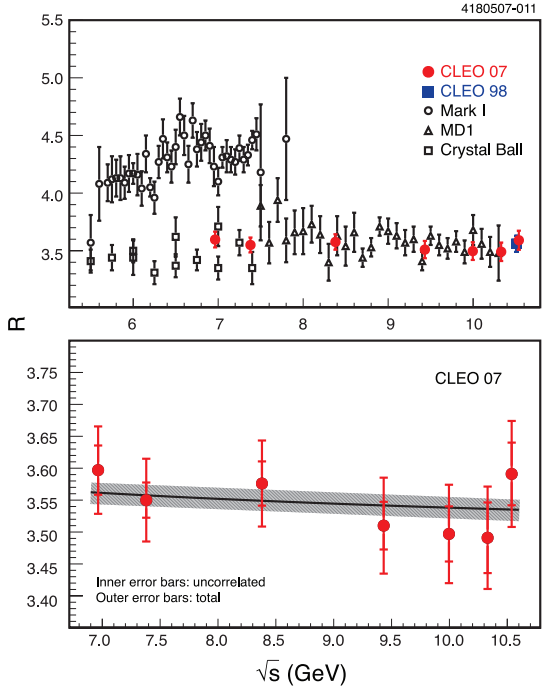}
\vspace{-1cm}
\caption{The upper figure compares the measurements of $R$ presented in this paper with previous 
measurements. The lower plot compares the results with predicted values of $R$; the width of band 
is given by the uncorrelated uncertainties on $\Lambda$.}
\label{fig:CLEOIII_R}
\end{figure}

All measurements presented are made with data collected at the Cornell Electron Storage Ring 
(CESR). The CLEO III detector \cite{bib:cleoIII} is used to study the data with 
$\sqrt{s}=6.964-10.538~\mathrm{GeV}$. The CLEO-c detector \cite{bib:cleoc} is used to study 
the data around $c\overline{c}$ threshold. The principal differences between the 
CLEO III and CLEO-c detectors are the reduction in solenoid field from 1.5 to 1.0~T and the 
replacement of CLEO III's four-layer silicon-strip detector by a six-layer all-stereo inner 
drift chamber. These modifications improve the reconstruction of low 
momentum charged particles.
 
\section{Measurements of $R$ at $\sqrt{s}=6.964-10.538~\mathrm{GeV}$}
\label{sec:cleoIII_R}
\begin{table}[t]
\caption{The centre-of-mass energy, luminosity and the value of $R$ for the seven 
measurement points with $\sqrt{s}=6.964-10.538~\mathrm{GeV}$. The combined statistical and 
systematic uncertainty is given on the luminosity. The first, second and third uncertainties 
on $R$ are the statistical, common systematic and uncorrelated systematic, respectively.}
\label{tab:cleoIII-res}
\newcommand{\m}{\hphantom{$-$}}
\newcommand{\cc}[1]{\multicolumn{1}{c}{#1}}
\begin{center}
\footnotesize
\begin{tabular}{@{}lll}
\hline
$\sqrt{s}$ (GeV)          & $\int\mathcal{L}dt~(\mathrm{pb}^{-1})$ & $R$ \\
\hline
10.538 & $904.50\pm 9.00$ & $3.591 \pm 0.003 \pm 0.067 \pm 0.049$\\
10.330 & $149.80\pm 1.60$ & $3.491 \pm 0.006 \pm 0.058 \pm 0.055$\\
9.996  & $432.60\pm 4.80$ & $3.497 \pm 0.004 \pm 0.064 \pm 0.043$\\
9.432  & $183.00\pm 2.00$ & $3.510 \pm 0.005 \pm 0.066 \pm 0.037$\\
8.380  & $  6.78\pm 0.06$ & $3.576 \pm 0.024 \pm 0.058 \pm 0.025$\\
7.380  & $  8.48\pm 0.07$ & $3.550 \pm 0.019 \pm 0.058 \pm 0.020$\\
6.964  & $  2.52\pm 0.02$ & $3.597 \pm 0.033 \pm 0.057 \pm 0.020$\\
\hline
\end{tabular}
\end{center}
\end{table}
The values of $\sqrt{s}$ and integrated luminosities, $\int\mathcal{L}dt$, for each data point 
are given in Table~\ref{tab:cleoIII-res}. Hadronic events are 
selected from these data by placing criteria on individual tracks and showers, as well as the 
whole event. 

 Tracks used in the analysis are required to be of good quality, originate from close to 
the interaction point and have a momentum between 1 and 150\% of the beam momentum. Showers must 
have at least 1\% of the beam momentum and not be 
associated with a track.

 Several event variables are used to discriminate against background.  The average point of 
origin of all tracks along the beam direction is used  to reject beam-gas, beam-wall and 
cosmic ray events. The total visible energy normalised to twice the beam energy and the missing 
momentum in the beam direction normalised to the visible energy are used to 
remove two-photon and beam-gas events. Each event must contain at least four charged tracks and 
pass an event shape criteria to reject 
$e^{+}e^{-}\rightarrow l^{+}l^{-}~(l=e,\mu~\mathrm{or}~\tau)$ events. The ratio of 
total calorimeter energy associated to tracks and isolated showers normalised to twice the 
beam energy is used to reject Bhabha and tau-pair events. Initial state radiation events are 
removed by a criterion on the ratio of the maximum isolated shower energy normalised to the beam 
energy. 

The efficiency of the selection increases from 82.1\% at $\sqrt{s}=6.964~\mathrm{GeV}$ to 87.4\% 
at $\sqrt{s}=10.538$. The dominant background remaining after the event selection is from 
$e^{+}e^{-}\rightarrow \tau^{+}\tau^{-}$ events, which is estimated 
from simulation. Other remaining backgrounds are estimated to give contributions smaller 
than the systematic uncertainty assigned to the event selection.

 Radiative corrections must be applied to the measured total hadronic cross section to determine 
$R$. 
Corrections are applied for soft 
photon radiation and vacuum polarisation. In addition, corrections are 
applied for hard initial state radiation to the continuum and lower mass $q\bar{q}$ 
resonances. Furthermore, at three energy points, a correction is applied for the interference 
between 
a nearby $\Upsilon$ resonance and the continuum.

  Several sources of significant systematic uncertainty are considered: luminosity, radiative 
corrections, multiplicity corrections and event selection criteria. 
The relative uncertainty on the luminosity is between 0.9\% and 1.1\% depending on $\sqrt{s}$. The 
uncertainty from the radiative 
corrections is dominated by those for the hadronic vacuum polarisation, which has an 
uncertainty of 1.0\%. Some disagreement is found comparing the charged track multiplicity 
distribution in data and simulation. Therefore, the efficiency is determined as a function 
of multiplicity before applying to the data. This procedure is estimated to have an 
uncertainty between 0.4\% and 1.4\% depending on $\sqrt{s}$. The uncertainty related to other 
selection criteria is estimated to be between 1.0 and 1.4\% depending on $\sqrt{s}$.

 The measured values of $R$ are given in Table~\ref{tab:cleoIII-res} along with the associated 
uncertainties. The total relative systematic uncertainty at 
each point is between 1.7 and 2.3\%. The measured values are compared to previous 
measurements \cite{bib:CLEOIII_previousR} in the upper plot in Figure~\ref{fig:CLEOIII_R}. 
The measurements are in agreement with the results from Crystal Ball, MD1 and CLEO; however, they 
do not agree with the MARK I results.

 The value of $\alpha_{s}$ is determined at each point using a perturbative QCD calculation of $R$ 
at the four-loop level \cite{bib:alphas_theory}. (This calculation ignores the quark masses; a 
determination including quark mass effects is presented in Ref. \cite{bib:KuhnT}.) The 
compatibility of these measurements with others of $\alpha_s$ is evaluated by exploiting the 
expected running of $\alpha_s$ \cite{bib:pdg2004}, which depends on the QCD scale $\Lambda$. The 
measured values determine $\Lambda = 0.31^{+0.09+0.29}_{-0.08-0.21}~\mathrm{GeV}$ and  
$\alpha(M^{2}_{Z})=0.126\pm 0.005^{+0.015}_{-0.011}$, where the first uncertainties are statistical 
and the second systematic. These results agree with the world averages \cite{bib:alphas_average}. 
The lower plot in Figure~\ref{fig:CLEOIII_R} compares the measured values of $R$ to those predicted 
by the fitted value of $\Lambda$.

\section{Studies of exclusive charm and total cross sections at $\sqrt{s}=3.97-4.26~\mathrm{GeV}$}

Data collected at thirteen energy points between $\sqrt{s}=3.97-4.26~\mathrm{GeV}$ are studied. 
\label{sec:cleoc_R}
\begin{figure}[h]
\includegraphics*[width=0.95\columnwidth]{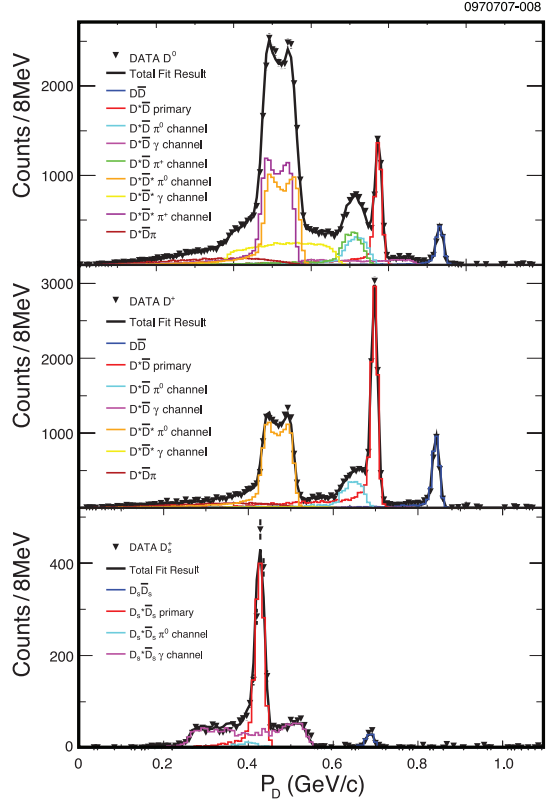}
\vspace{-1cm}
\caption{The sideband-subtracted momentum spectra for $D^{0}$ (upper), $D^{+}$ (middle) and $D_{s}$ 
(lower) for data collected at $\sqrt{s}=4.17~\mathrm{GeV}$. The fit results for the different 
production mechanisms are also shown.}
\label{fig:CLEOc_Dmom}
\end{figure}
\begin{figure}[h]
\includegraphics*[width=0.95\columnwidth]{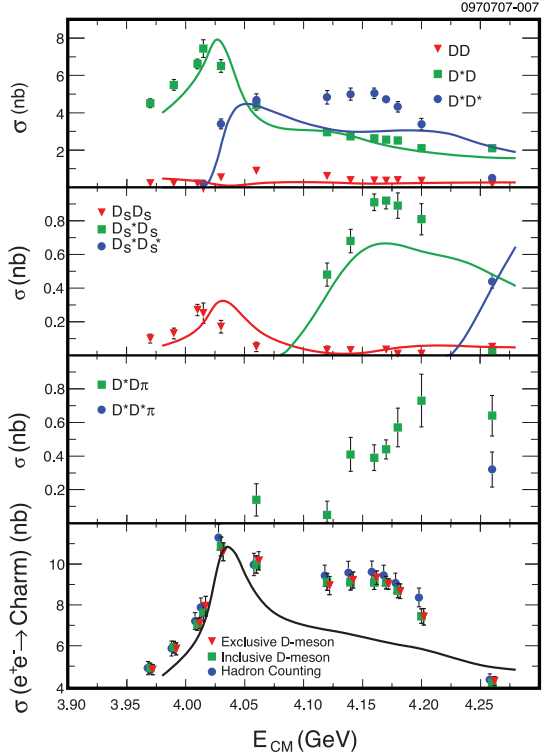}
\vspace{-1cm}
\caption{The production cross sections for $D^{(*)}\overline{D^{(*)}}$ (upper), $D^{(*)}_s 
\overline{D^{(*)}_s}$ (upper middle) and multi-body (lower middle) production. The lower plot shows 
the total charm cross section determined by the alternate methods described in the text. The 
two-body and total production cross sections are compared to a model \cite{bib:eichten}.} 
\label{fig:CLEOc_charmprod}
\end{figure}
Specific criteria are used to select $D^{+}$, $D^{0}$ and $D_{s}^{+}$ candidates; these closely 
follow other CLEO-c analyses \cite{bib:cleoc_dhad}. The final states with the largest 
signal-to-background ratio are considered for $D^{+}\rightarrow K^{-}\pi^{+}\pi^{+}$ and 
$D^{0}\rightarrow K^{-}\pi^{+}$. Eight decay modes are reconstructed to select $D^{+}_{s}$ 
candidates: $\phi(K^{+}K^{-})\pi^{+}$, $K^{*0}(K^{-}\pi^{+})K^{+}$, $\eta(\gamma\gamma)\pi^{+}$, 
$\eta(\gamma\gamma)\rho^{+}(\pi^{+}\pi^{0})$, 
$\eta^{\prime}(\pi^{+}\pi^{-}\eta(\gamma\gamma))\pi^{+}$, 
$\eta^{\prime}(\pi^{+}\pi^{-}\eta(\gamma\gamma))\rho^{+}(\pi^{+}\pi^{0})$, 
$\phi(K^{+}K^{-})\rho^{+}(\pi^{+}\pi^{0})$ and $K^{0}_{S}(\pi^{+}\pi^{-})K^{+}$. These modes 
correspond to 16\% of the total $D^{+}_{s}$ decay width.

 Candidates are selected if their mass lies within $\pm 15~\mathrm{MeV}$ of the nominal $D_{(s)}$ 
values. Background is subtracted using yields measured in $D_{(s)}$ mass sidebands extrapolated 
into the signal region. The momentum distribution of the $D_{(s)}$ candidates is then used to 
determine the production mechanism. The $D^{0}$, $D^{+}$ and $D^{+}_{s}\rightarrow\phi\pi^{+}$ 
momentum distribution at 
$\sqrt{s}=4.17~\mathrm{GeV}$ is shown in Figure~\ref{fig:CLEOc_Dmom}. The different peaks 
correspond to different production mechanisms. The shape of distributions for the individual 
processes are determined from simulation. The data are fit to the different components to determine 
the yields. For $D^{+}$ and $D^{0}$ production these are then corrected for efficiency, branching 
ratios and luminosity to give the cross sections shown in the two upper plots in 
Figure~\ref{fig:CLEOc_charmprod}. Two additional components describing multi-body production 
$D^{*}\overline{D^{(*)}}\pi$ are required to describe the low momentum distribution at 
$\sqrt{s}>4.06~\mathrm{GeV}$. This is the first observation of multi-body production in the charm 
threshold region.
     
Given the relative simplicity of the $D^{(*)+}_{s} D^{(*)-}_{s}$ production and the limited 
statistics an alternative technique is used to determine the cross sections. The separation of the 
different mechanisms in the beam-energy difference $(\Delta E = E_{D_{s}} - E_{beam})$ and the 
beam-constrained mass $(M_{bc} = \sqrt{E^{2}_{beam}-|\mathbf{P}_{D_{s}}|^{2}})$ plane is used. The 
background is subtracted using $\Delta E$ and $M_{bc}$ sidebands. The resulting cross sections are 
shown in the upper middle plot in Figure~\ref{fig:CLEOc_charmprod}. 

The dominant sources of systematic uncertainty are the selection efficiency, yield determination 
and the normalisation. Details of the uncertainty on the selection efficiency can be found in 
Ref.~\cite{bib:cleoc_dhad}. The signal functions for the determination of $D^{0}$ and $D^{+}$ 
production mechanisms depend on the modelling of initial state radiation and the helicity 
amplitudes for $D^{*}D^{*}$; variations of these models over a broad range of assumptions leads to 
the systematic uncertainty. Variations in the $M_{bc}$ and $\Delta E$ selection criteria are used 
to estimate the uncertainty related to signal extraction in the $D_{s}$ modes. The uncertainty on 
the normalisation arises from that on the measured luminosity and the branching fractions of the 
modes reconstructed \cite{bib:pdg2006}. The total systematic uncertainties are between 3.4\% and 
6.8\% for two-body production mechanisms; the multi-body production mechanisms 
$D^{*}\overline{D}\pi$ and $D^{*}\overline{D}^{*}\pi$ have systematic uncertainties of 12\% and 
25\% uncertainties, respectively.       

The results are compared to an updated calculation of Eichten {\it et al.} \cite{bib:eichten}. 
There is reasonable qualitative agreement for most two-body production mechanisms apart from 
$D^{*}\overline{D^{*}}$ production in the $\sqrt{s}$ range 4.05 to 4.20~GeV. 

The results at $4.26~\mathrm{GeV}$ have the potential to study the nature of the $Y(4260)$. Hybrid 
charmonium \cite{bib:hybrid_charm} and tetraquark \cite{bib:tetraquark} interpretations suggest 
enhancements of some production mechanisms; no significant enhancements are observed disfavouring 
these models.

The sum of the exclusive cross sections should equal the total charm cross section. This has been 
tested with measurements using two inclusive techniques. 
The sum of inclusive cross-sections for single $D^{0}$, $D^{+}$ and $D_{s}$ production 
divided by two is found to be in agreement with the total exclusive cross sections. In addition, 
the total hadronic cross section is determined in a manner similar to that described in 
Section~\ref{sec:cleoIII_R}. The light-quark production cross section is subtracted using 
measurements below $c\bar{c}$ threshold extrapolated with $1/s$ dependence. The total charm cross 
section from this method is found to be in agreement with the other two techniques. The cross 
sections determined by each method are compared to each other and a model in the lower plot of 
Figure~\ref{fig:CLEOc_charmprod}.

\begin{figure}[htb]
\vspace{-1cm}
\includegraphics*[width=0.95\columnwidth]{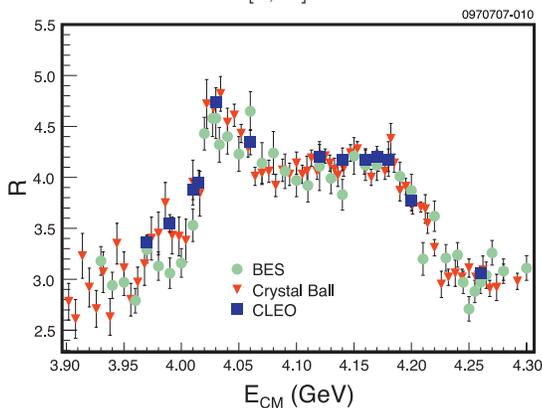}
\vspace{-1cm}
\caption{Measurements of $R$ at thirteen points in the $c\bar{c}$ threshold region.}
\label{fig:CLEOc_R}
\end{figure}

The total hadronic cross section is radiatively corrected \cite{bib:kuraev_and_fadin} to obtain the 
measurements of $R$, which is shown in Figure~\ref{fig:CLEOc_R}. These measurements are more 
precise and in good agreement with previous measurements \cite{bib:bes_R,bib:CB_R}.   

\section{Conclusions}
\label{sec:conclusions}
Seven measurements of $R$ in the range $\sqrt{s}=6.964-10.538~\mathrm{GeV}$ are used to determine a 
value of $\alpha_s(M_{Z}^{2})$ in agreement with the world average. The exclusive and inclusive 
charm cross sections are measured at thirteen $\sqrt{s}$ values near $c\bar{c}$ threshold. 
Multi-body production $D^{*}\overline{D^{(*)}}\pi$ is observed for the first time. All measurements 
of $R$ are more precise and in agreement with earlier measurements.

\end{document}